\documentclass[letterpaper, 10 pt, conference]{ieeeconf}  
\IEEEoverridecommandlockouts
\usepackage{cite}
\usepackage{amsmath,amssymb,amsfonts}
\usepackage{bm}
\usepackage{algorithmic}
\usepackage{graphicx}
\usepackage{textcomp}
\usepackage{xcolor}
\usepackage{graphicx}
\usepackage{epsfig}
\usepackage{epstopdf}
\usepackage{mathptmx}
\usepackage{times}
\usepackage{bm}
\usepackage{stackengine}
\def\delequal{\mathrel{\ensurestackMath{\stackon[1pt]{=}{\scriptstyle\Delta}}}}
\usepackage{url}
\usepackage{flushend}
\usepackage{multirow}
\usepackage{array}
\usepackage{color}
\usepackage{cite}

\usepackage{ntheorem}

\usepackage{proof-at-the-end}
\usepackage{algorithm}
\usepackage{algorithmic}      
\usepackage{multirow}
\pgfkeys{/prAtEnd/global custom defaults/.style={
one big link={See proof in Section~\ref{proofssection}.}
}
}

\usepackage{graphics} 
\usepackage{epsfig} 
\usepackage{mathptmx} 
\usepackage{times} 
\usepackage{color}
\usepackage{cite}
\usepackage{graphicx}
\usepackage{subfigure}
\usepackage{multicol}

\usepackage{xparse}
\NewDocumentEnvironment{mynormalthm}{O{}O{}+b}{%
  \begin{theoremEnd}[normal,#2]{thm}[#1]%
    #3%
  \end{theoremEnd}%
}{}

\pgfkeys{/prAtEnd/global custom defaults/.style={
one big link={See proof in Section~\ref{proofs}.}
}
} 
\newtheorem{theorem}{Theorem}[section]
\newtheorem{lemma}[theorem]{Lemma}

\newtheorem{assumption}[theorem]{Assumption}

\newtheorem{definition}[theorem]{Definition}

\newtheorem{remark}[theorem]{Remark}

\def\BibTeX{{\rm B\kern-.05em{\sc i\kern-.025em b}\kern-.08em
    T\kern-.1667em\lower.7ex\hbox{E}\kern-.125emX}}
\begin{document}

\title{Dissensus Algorithms for Opinion Dynamics on the Sphere \\
}
\author{Ziqiao Zhang, Said Al-Abri, and Fumin Zhang
\thanks{Ziqiao Zhang, Said Al-Abri,  and Fumin Zhang are at School of Electrical and Computer
Engineering, Georgia Institute of Technology, Atlanta, GA 30332, E-mail:
{\tt\small \{ziqiao.zhang,saidalabri,fumin\}@gatech.edu}.
}%
}
\maketitle

\begin{abstract}
In this paper, novel dissensus algorithms based on the Oja principal component analysis (PCA) flow are proposed to model opinion dynamics on the unit sphere. 
The information of the covariance formed by the opinion state of each agent is used to achieve a dissensus equilibrium on unsigned graphs. This differs from most of the existing work where antagonistic interactions represented by negative weights in signed graphs are used to achieve a dissensus equilibrium. The nonlinear algorithm is analyzed under both constant covariance and time-varying covariance leading to different behaviors.  
Stability analysis for the unstable consensus and stable dissensus equilibria is provided under various conditions. The performance of the algorithm is illustrated through a simulation experiment of a multi-agent system.
\end{abstract}


\section{Introduction}

In social networks, opinion dynamics have been studied to understand not only how individuals exchange opinions with their neighbors and form their own opinions during the process of information exchange, but also group behaviors \cite{PROSKURNIKOV201765}. While the opinions evolve within the group, it is natural that the individuals tend to reach consensus or dissensus gradually. In this way, a group decision can be made, whether they agree or disagree on a certain topic. Consensus/dissensus algorithms have been applied in distributed source seeking \cite{8263865}, cooperative control of multi-agent system \cite{6645381,SARLETTE2009572}, opinion-forming in social networks \cite{6866175,golub2010naive}.


For connected and undirected graphs, there is a class of averaged consensus protocols that yield almost global consensus on the unit sphere \cite{markdahl2017almost}. Using the same averaged consensus algorithm but allowing negative weights for graph connections,  stable dissensus on the unit sphere can also be achieved if the interaction matrix is sign-symmetric  \cite{caponigro2015nonlinear}. In \cite{Altafini2013}, dissensus is described as \textit{bipartite consensus}, and linear/nonlinear consensus protocols can be applied to achieve stable dissensus on a signed graph with negative weights representing antagonistic relationships and positive weights representing collaborative relationships. In \cite{Ma2018}, a distributed algorithm is developed to reach dissensus for a multi-agent system on a signed digraph. These dissensus behaviors mostly appear in signed graphs, but not in unsigned graphs.

In this paper, we propose an Oja principal component analysis (PCA) based dissensus algorithm for the nonlinear opinion dynamics evolved on the surface of the unit sphere. Given a group of agents, each agent forms its own opinion according to the nonlinear dynamics on the unit sphere.  
The interaction among agents is modeled by the covariance matrix of the relative opinions of its neighbors. By combining the nonlinear dynamics on the unit sphere with the covariance-based interaction, we set up the Oja PCA opinion dynamics and show that the opinion states of the agents in the group will reach stable dissensus equilibrium under certain conditions. The novel contribution of this paper is the application of  Oja PCA flow to model opinion dynamics in multi-agent systems. Instead of using averaged consensus algorithm, we are using a PCA-based method which is based on the real-time difference between individuals. Under this PCA-based opinion dynamics, the opinion system is able to achieve a stable dissensus equilibrium starting from non-equilibrium initial conditions. In \cite{Altafini2013}, the stable dissensus is driven by the antagonistic information exchange within the group. In our case, the covariance matrix describes the opinion differences, and such difference is maximized when the opinions are at dissensus. The differences between individual opinions lead to stable dissensus.
Unlike the dissensus algorithms based on signed graphs, our approach achieves stable dissensus for unsigned graphs. Hence no extra information about collaborative or antagonistic interactions between individuals is needed.  

The main challenge the paper has overcome is the convergence analysis of Oja PCA flow with time-varying covariance. For Oja PCA flow with fixed matrix, \cite{oja1982simplified}, it has been shown that the solutions of Oja's equation will converge to the principal eigenspace of the fixed matrix \cite{317720,yoshizawa2001convergence}. However, for the time-varying matrix case, the convergence of Oja PCA flow has not been considered before. In this paper,  Oja PCA flow has been used to model opinion dynamics and the fixed matrix has been replaced by the time-varying covariance of the opinion states. A special case of Oja PCA flow with the time-varying matrix is constructed in this way, and the corresponding convergence analysis is given by a stability proof of dissensus equilibrium under the PCA-based opinion dynamics. This is difficult in that both the opinion dynamics and the dynamics of the covariance matrix need to be analyzed, and the two types of dynamics make the stability analysis non-trivial.

The main contributions of this paper are as follows. The first contribution is proposing novel modeling of opinion dynamics on the unit sphere using an Oja PCA flow. The second contribution is using a time-varying covariance of the opinion states to achieve stable dissensus. The third contribution is providing various stability results. In particular, (i) we prove via Lyapunov analysis that the consensus equilibrium in an $N-$agent network is unstable, (ii) we prove via linearization that the dissensus equilibrium in an $N-$agent network is stable, and (iii) we derive the region of attraction of the nonlinear system for the $2-$agent and $3-$agent networks. The final contribution is illustrating the behavior of the algorithm through a simulation experiment of a 20-agent system in $\mathbb{R}^2$.

The rest of the paper is organized as follows. The problem formulation is given in Section~\ref{problemformulation}. Then, the design and convergence analysis of the nonlinear opinion dynamics with constant and time-varying covariance matrices  is presented in Section~\ref{PCAdirect} and Section~\ref{PCAmultiagent}, respectively. Finally, simulation experiments are presented in Section~\ref{simulation} and concluding remarks are provided in Section~\ref{conclusion}. To increase the readability of the paper, some proofs are given in the Appendix in Section~\ref{proofs}.


\section{Problem Formulation}\label{problemformulation}
Consider a group of $N$ agents exchanging opinions about certain options where $N\in\mathbb{N}$ and $N\geq 2$. The interactions between the agents in the group are described by a graph $\mathcal{G}=(\mathcal{V},\mathcal{E})$ where $\mathcal{V}$ is the set of all agents and $\mathcal{E}$ is the set of all edges.

\begin{assumption}\label{graph_complete}
The graph $\mathcal{G}=(\mathcal{V},\mathcal{E})$ formed by the group of agents is undirected and completely connected, i.e. $(i,j)\in\mathcal{E}$ and $(j,i)\in\mathcal{E}$ for any $i,j\in\mathcal{V}, i\ne j$.
\end{assumption} 

\begin{assumption}\label{graph_unsigned}
The graph $\mathcal{G}=(\mathcal{V},\mathcal{E})$ formed by the group of agents is unweighted and unsigned, i.e. the opinion states from other agents are viewed with equal weight of $1$ for each agent $i$ and there is no antagonistic interaction between agents described by the graph.
\end{assumption} 

\begin{remark}
Different from unsigned graph, there are negative weights if a graph is signed. Negative weights are used to indicate competitive relationships between agents.
\end{remark}

The opinion of any agent $i\in\mathcal{V}$ is represented by a unit-length vector $\bm{v}_i\in\mathbb{R}^d$, $\lVert\bm{v}_i\rVert_2=1$. Let $\bm{v}=[\bm{v}_1,\cdots,\bm{v}_N]^\intercal$ be a vector containing opinions states of all agents.
Each opinion state evolves on the surface of the unit sphere $\mathbb{S}^{d-1}$ according to the nonlinear dynamics
\begin{align}\label{odynamics}
    \Dot{\bm{v}}_i = (\bm{I}-\bm{v}_i\bm{v}_i^\intercal)\bm{u}_i
\end{align}
where $\bm{I}\in\mathbb{R}^{d\times d}$ is the identity matrix and $\bm{u}_i=\bm{u}_i(\bm{v})\in\mathbb{R}^d$ is a control input for agent $i$. 

\begin{definition} [Consensus Behavior]\label{def_con}
The opinion states of the agents in the group are in consensus if 
\begin{align}
    \bm{v}_i=\bm{v}_j,\,\,\forall (i,j)\in\mathcal{E}, 
\end{align}
where $\mathcal{E}$ is the set of all edges of the graph.
\end{definition}

\begin{definition} [Dissensus Behavior]\label{def_dis}
The opinion states of the agents in the group are in dissensus if there exist two non-empty sets $\mathcal{V}_1,\mathcal{V}_2$ satisfying
\begin{align}
    \mathcal{V}_1\cup\mathcal{V}_2=\mathcal{V}, \mathcal{V}_1\cap\mathcal{V}_2=\emptyset,
\end{align}
such that
\begin{align}
    \bm{v}_i=-\bm{v}_j,\quad\forall (i,j)\in\bar{\mathcal{E}},
\end{align}
where $\bar{\mathcal{E}}\delequal \{(i,j)|i\in\mathcal{V}_1\text{ and }j\in\mathcal{V}_2\}$ is the set of all edges connecting agents in sets $\mathcal{V}_1$ and $\mathcal{V}_2$.
\end{definition}

Let $\bm{C}(\bm{v})\in\mathbb{R}^{d\times d}$ be a positive semi-definite covariance matrix defined by
\begin{align}\label{cov}
    \bm{C}(\bm{v})&=\sum_{k\in\mathcal{V}}(\bm{v}_k(t)-\Bar{\bm{v}}(t))(\bm{v}_k(t)-\Bar{\bm{v}}(t))^\intercal,
\end{align}
where $\Bar{\bm{v}}(t)=\frac{1}{|\mathcal{V}|}\sum_{i\in\mathcal{V}}\bm{v}_i(t)$ is the average of all opinions.

The opinion dynamics modeled by the Oja PCA flow under a constant covariance $\bm{H}=\bm{C}(\bm{v}(0))$ is given by
\begin{align}\label{pca_dynamics_constant_H}
    \Dot{\bm{v}}_i = (\bm{I}-\bm{v}_i\bm{v}_i^\intercal)\bm{H}\bm{v}_i,\quad\forall i\in\mathcal{V},
\end{align}
where $\bm{H}=\bm{C}(\bm{v}(0))=\sum_{k\in\mathcal{V}}(\bm{v}_k(0)-\Bar{\bm{v}}(0))(\bm{v}_k(0)-\Bar{\bm{v}}(0))^\intercal$ is a fixed covariance matrix defined based on the initial opinions. In other words, in (\ref{pca_dynamics_constant_H}) all agents exchange their opinion states $\bm{v}_i(t)$ only at time $t=0$. 

Alternatively, let the control input $\bm{u}_i$ to be
\begin{align}\label{pcacontrol}
    \bm{u}_i=\sum_{k\in\mathcal{V}}\langle \bm{v}_k-\Bar{\bm{v}},\bm{v}_i\rangle(\bm{v}_k-\Bar{\bm{v}})=
    \bm{C}(\bm{v})\bm{v}_i,
\end{align}
which by substituting \eqref{pcacontrol} into  \eqref{odynamics} leads to the opinion dynamics
\begin{align}\label{pca_dynamics}
    \Dot{\bm{v}}_i = (\bm{I}-\bm{v}_i\bm{v}_i^\intercal)\bm{C}(\bm{v})\bm{v}_i,\quad \forall i\in\mathcal{V}.
\end{align}
Note that, in contrast to (\ref{pca_dynamics_constant_H}), the agents under the opinion dynamics (\ref{pca_dynamics}) need to exchange their opinion state $\bm{v}_i(t)$ at each instant of time $t$.

In this paper, we aim to study the the behavior of the Oja PCA opinion dynamics under both a constant covariance matrix $\bm{H}$ in \eqref{pca_dynamics_constant_H},  and a varying covariance matrix $\bm{C}(\bm{v})$ in \eqref{pca_dynamics}, respectively. In particular, our purpose is to derive the equilibrium points of \eqref{pca_dynamics_constant_H} and \eqref{pca_dynamics}, and determine the conditions under which the system will achieve either a consensus or a dissensus behavior. Moreover, we aim to compare our PCA-based modeling of opinion dynamics to the conventional average-based  opinion dynamics where $\bm{u}_i=\sum_{k\in\mathcal{V}}\bm{v}_k$ \cite{caponigro2015nonlinear}.

\begin{remark}
To the best of our knowledge, the Oja PCA flow  has never been used before to model opinion dynamics. Additionally, the existing convergence analysis and results of the Oja PCA flow, e.g., \cite{oja1982simplified,317720,yoshizawa2001convergence}, do not hold for the  opinion dynamics \eqref{pca_dynamics}. This is due to the fact that the matrix $\bm{C}(\bm{v})$ in our modeling is time-varying, while it is constant in existing works \cite{317720,yoshizawa2001convergence}.
\end{remark}
\begin{remark}
The objective of using a PCA-based formulation for opinion dynamics is to achieve a stable dissensus equilibrium on unsigned graphs. This cannot be achieved by conventional consensus-on-a sphere algorithms where they require signed graphs to achieve a stable dissensus.   
\end{remark}

\section{PCA-Based Opinion Dynamics with Constant Covariance
}\label{PCAdirect}
In this section,  we model the opinion dynamics of an $N$-agent system using an Oja PCA flow with a general constant matrix $\bm{H}$. We analyze the convergence of the system based on the initial opinion states. Based on this analysis, we propose a mechanism  to design an arbitrary constant matrix $\bm{H}$, based on a  given initial conditions $\{\bm{v}_k(0)\}_{k\in\mathcal{V}}$, that will yield a steady-state consensus or dissensus behavior.

\begin{assumption}\label{constant_H}
The constant matrix $\bm{H}$ is positive semi-definite. And the largest eigenvalue $\lambda_1$ of $\bm{H}$ is strictly positive and has multiplicity 1.
\end{assumption}

Let $\bm{q}$ be the unit eigenvector of $\bm{H}$ corresponding to the largest eigenvalue $\lambda_1$, i.e. $\bm{H}\bm{q}=\lambda_1\bm{q}$. We first analyze the dynamics of each $\bm{v}_i$ separately.

\begin{lemma}\label{q_converge}
Suppose each agent updates its opinion state according to  the Oja PCA dynamics \eqref{pca_dynamics_constant_H} with constant $\bm{H}$.
Then, for each $i\in\mathcal{V}$, $\bm{v}_i$ is converging to $\bm{q}$ if $0<\bm{q}^\intercal\bm{v}_i(0)\leq1$, and is converging to $-\bm{q}$ if $-1\leq\bm{q}^\intercal\bm{v}_i(0)<0$ for any $i\in\mathcal{V}$. Otherwise, if $\bm{q}^\intercal\bm{v}_i(0)=0$, then $\bm{v}_i$ remains unchanged.
\end{lemma}
\begin{proof}
To prove that \eqref{pca_dynamics_constant_H} converges to $\bm{q}$ if $0<\bm{q}^\intercal\bm{v}_i(0)\leq1$, we define  $\beta_i=1-\bm{q}^\intercal\bm{v}_i$ where $\beta_i=0$ if and only if $\bm{v}_i=\bm{q}$. Then we obtain
\begin{align}\label{betaidot}
    \dot{\beta}_i=-\bm{q}^\intercal\dot{\bm{v}}_i&=-\bm{q}^\intercal(\bm{I}-\bm{v}_i\bm{v}_i^\intercal)\bm{H}\bm{v}_i\notag\\
    &=-\bm{q}^\intercal\bm{H}\bm{v}_i+\bm{q}^\intercal\bm{v}_i\bm{v}_i^\intercal\bm{H}\bm{v}_i\notag\\
    &=-\lambda_1\bm{q}^\intercal\bm{v}_i+\bm{q}^\intercal\bm{v}_i\bm{v}_i^\intercal\bm{H}\bm{v}_i\notag\\
    &=(1-\beta_i)(\bm{v}_i^\intercal\bm{H}\bm{v}_i-\lambda_1).
\end{align}
Define a Lyapunov candidate function $V=\beta_i$ which implies that $\dot{V}=\dot{\beta}_i$ where $\dot{\beta}_i$ is as given by \eqref{betaidot}.
If $0<\bm{q}^\intercal\bm{v}_i(0)<1$, then $\beta_i(0)\in[0,1)$. Additionally, since $\lambda_1$ is the largest eigenvalue of $\bm{H}$, then at any time $\bm{v}_i^\intercal\bm{H}\bm{v}_i\leq\lambda_1$ and $\bm{v}_i^\intercal\bm{H}\bm{v}_i=\lambda_1$ if and only if $\beta_i=0$. Hence, $\dot{V}(0)\leq0$ and $\dot{V}(0)=0$ if and only if $\beta_i=0$. Since $V=\beta_i$ is a monotonic function and bounded below by zero, then the trajectory of \eqref{betaidot} will stay in a compact sublevel set of the Lyapunov function, implying that $0<\bm{q}^\intercal\bm{v}_i(t)<1$, or  $\beta_i(t)\in[0,1)$, for all $t>0$. Therefore, the origin $\beta_i=0$ 
 is asymptotically stable and hence $\bm{v}_i\to\bm{q}$ as $t\to\infty$.
 
 On the other hand, to  \eqref{pca_dynamics_constant_H} converges to $-\bm{q}$ if $-1\leq\bm{q}^\intercal\bm{v}_i(0)<0$, we define  $\alpha_i=1+\bm{q}^\intercal\bm{v}_i$ where $\alpha_i=0$ if and only if $\bm{v}_i=-\bm{q}$. Then, similar to \eqref{betaidot} we obtain
\begin{align}\label{alphaidot}
    \dot{\alpha}_i=\bm{q}^\intercal\dot{\bm{v}}_i=(1-\alpha_i)(\bm{v}_i^\intercal\bm{H}\bm{v}_i-\lambda_1).
\end{align}
Since \eqref{betaidot} and \eqref{alphaidot} are equivalent, then we can also conclude that the origin $\alpha_i=0$ of \eqref{alphaidot}
 is asymptotically stable and hence $\bm{v}_i\to-\bm{q}$ as $t\to\infty$ whenever $-1\leq\bm{q}^\intercal\bm{v}_i(0)<0$. Finally, if $\bm{q}^\intercal\bm{v}_i(0)=0$, then $\dot{\bm{v}}_i=0$, which is another equilibrium. Since the system \eqref{pca_dynamics_constant_H} is assumed to be noiseless, then $\bm{q}^\intercal\bm{v}_i(t)=0$ for all $t>0$.
\end{proof}
Since $\bm{H}$ in \eqref{pca_dynamics_constant_H} is constant,  $\dot{\bm{v}}_i$ is only determined by $\bm{v}_i(0)$ and $\bm{H}$. Additionally,  since there is no information exchange between agents, then this implies that whether $\{\bm{v}_i\}_{i\in\mathcal{V}}$ reaches consensus or dissensus mainly depends on the initial opinions $\{\bm{v}_i(0)\}_{i\in\mathcal{V}}$ and $\bm{H}$.

For given initial conditions, we can design constant matrix $\bm{H}$ to achieve consensus or dissensus as follows
\begin{itemize}
    \item If there exists $\bm{q}\in\mathbb{R}^d$ and $\|\bm{q}\|_2=1$ such that $0<\bm{q}^\intercal\bm{v}_i(0)\leq1$ holds for any $i\in\mathcal{V}$, then we can find a constant matrix $\bm{H}=\bm{q}\bm{q}^\intercal$ to achieve consensus equilibrium of $\bm{q}$. Such $\bm{H}$ satisfies Assumption \ref{constant_H} and $\bm{H}\bm{q}=\lambda_1\bm{q}$ with $\lambda_1=\bm{q}^\intercal\bm{q}=1$ being the largest eigenvalue. 
    \item If there exists $\bm{q}\in\mathbb{R}^d$, $\|\bm{q}\|_2=1$ and the initial opinion states can be divided into two groups $\mathcal{V}_1$ and $\mathcal{V}_2$, such that $0<\bm{q}^\intercal\bm{v}_i(0)\leq1$ for any $i\in\mathcal{V}_1$ and $-1\leq\bm{q}^\intercal\bm{v}_i(0)<0$ for any $j\in\mathcal{V}_2$, then the opinion states are converging to dissensus equilibrium of $\{\bm{q},-\bm{q}\}$ for constant $\bm{H}=\bm{q}\bm{q}^\intercal$ satisfying Assumption \ref{constant_H} and $\bm{H}\bm{q}=\lambda_1\bm{q}$ with $\lambda_1=\bm{q}^\intercal\bm{q}=1$ being the largest eigenvalue.
\end{itemize}

\begin{remark}
For some cases of initial conditions, it is possible to design different $\bm{H}$ to achieve consensus and dissensus separately. But there are cases that only consensus or only dissensus can be achieved. For example, a 2-agent system with $\bm{v}_1=-\bm{v}_2$ cannot achieve consensus for any $\bm{H}$, and a 2-agent system with $\bm{v}_1=\bm{v}_2$ cannot achieve dissensus for any $\bm{H}$.
\end{remark}
Once the initial conditions and $\bm{H}$ are fixed, whether $\bm{v}_i$ is converging to $\bm{q}$, $-\bm{q}$ or staying there can be determined according to Lemma \ref{q_converge}.
Therefore, in the case of constant $\bm{H}$, the opinion dynamics given by \eqref{pca_dynamics_constant_H} can provide stable consensus or stable dissensus equilibria depending on the initial conditions of opinion states and matrix $\bm{H}$. We cannot guarantee such algorithm always leads to dissensus for any constant $\bm{H}$ satisfying Assumption \ref{constant_H} and such algorithm cannot be served as a reliable dissensus algorithm for the general case.

Selecting a constant $\bm{H}$ requires that the matrix $\bm{H}$ be agreed by all agents, which can be achieved through averaged consensus. In the next section, we will introduce a new approach that does not use a constant matrix $\bm{H}$. Instead, the covariance matrix $\bm{C}(\bm{v})$ is used to replace $\bm{H}$ to achieve stable dissensus.


\section{PCA-Based Opinion Dynamics with Varying Covariance}\label{PCAmultiagent}
In this section, Oja PCA flow with a varying covariance matrix $\bm{C}(\bm{v})$ is applied to model opinion dynamics in $N$-agent system ($N\geq 2$). We will show that this   PCA-based opinion dynamics with a varying covariance will lead to unstable consensus and stable dissensus equilibria.  This is different from the opinion dynamics with constant $\bm{H}$ introduced in \eqref{pca_dynamics_constant_H}. 
\subsection{Unstable Consensus Equilibrium for $N-$agent Network}
\begin{theorem}\label{unstable_con_N}
Under the PCA dynamics \eqref{pca_dynamics}, the consensus equilibrium $\bm{v}_1=\bm{v}_2=\cdots=\bm{v}_N$ is unstable.
\end{theorem}
\begin{proof}
Define  $\beta_{i}=1-\langle\bm{v}_i,\bm{v}_N\rangle$, $i=1,\cdots,N-1$,  where $\beta_{i}\in[0,2]$, and  $\beta_{i}=0$ if and only if $\langle\bm{v}_i,\bm{v}_N\rangle=1$. Let $\beta=[\beta_1,\cdots,\beta_{N-1}]^\intercal$. 
Consider the Lyapunov candidate function 
\begin{align}\label{Vus1}
V(\beta)&=\sum_{i=1}^{N-1}\beta_i=
\frac12\sum_{i=1}^{N-1}\lVert \bm{v}_i-\bm{v}_N\rVert_2^2,
\end{align}
where $V\geq0$ and $V=0$ if and only if $\beta_i=1$, for $i=1,\cdots,N-1$, i.e.  $\bm{v}_1=\bm{v}_2=\cdots=\bm{v}_N$. Let $\Tilde{\beta}=[\epsilon,0,0,\cdots,0]^\intercal$ where $\epsilon>0$, i.e. $\bm{v}_2=\cdots=\bm{v}_{N-1}=\bm{v}_N$ while $\bm{v}_1\ne \bm{v}_N$. Hence, at $\beta=\Tilde{\beta}$, \eqref{Vus1} reduces to
\begin{align}
V(\Tilde{\beta})&=\beta_1=\frac12\lVert \bm{v}_1-\bm{v}_N\rVert_2^2>0.
\end{align}
Next, we obtain
\begin{align}
    \dot{V}(\Tilde{\beta})&=(\bm{v}_1-\bm{v}_N)^\intercal(\dot{\bm{v}}_1-\dot{\bm{v}}_N)\notag\\
 &=( \bm{v}_1-\bm{v}_N)^\intercal\left((\bm{I}-\bm{v}_1\bm{v}_1^\intercal)\bm{C}\bm{v}_1-(\bm{I}-\bm{v}_N\bm{v}_N^\intercal)\bm{C}\bm{v}_N\right).\notag
\end{align}
However, when $\beta=\Tilde{\beta}$
\begin{align}
\bm{C}=\sum_{i=1}^N(\bm{v}_i-\bar{\bm{v}})(\bm{v}_i-\bar{\bm{v}})^\intercal=\frac{N-1}{N}(\bm{v}_1-\bm{v}_N)(\bm{v}_1-\bm{v}_N)^\intercal.\notag
\end{align}
Hence,
\begin{align}
\dot{V}(\Tilde{\beta})&=\frac{2N-2}{N}(1-\bm{v}_1^\intercal\bm{v}_N)^2(1+\bm{v}_1^\intercal\bm{v}_N)
=\frac{2N-2}{N}\beta_1^2(2-\beta_1).\notag
\end{align}
Clearly, $\dot{V}(\Tilde{\beta})>0$ everywhere except at $\beta_1=0,2$, which is equivalent to
the equilibrium $\bm{v}_1=\pm\bm{v}_N$. Define the set $\bm{U}=\{\beta\in\bm{B}|V(\beta)>0\}$ where $\bm{B}=\{\beta\in\mathbb{R}|\beta<2(N-1)\}$. Note that the set $\bm{U}$ is nonempty set contained in $\bm{B}$. This implies that $\dot{V}(\beta)>0$ for all points in $\bm{U}$. Therefore, all the conditions in Theorem 4.3 in \cite{khalil2002nonlinear} are met, and hence the equilibrium $\beta=\bm{0}$, or equivalently $\bm{v}_1=\bm{v}_2=\cdots=\bm{v}_N$ is unstable.
\end{proof}
\begin{remark}
For multi-agent systems on unsigned graphs, classical consensus algorithms use (weighted) average of neighbors' opinions as a control input to achieve stable consensus on the unit sphere \cite{caponigro2015nonlinear}. The proposed PCA-based algorithm always leads to unstable consensus, and it can be a candidate of dissensus algorithm. In the next part, we will show this algorithm can give stable dissensus.
\end{remark}

\subsection{Linearization-based Stability Analysis of the Dissensus Equilibrium for $N-$agent Network}
In this section, we show via linearization that the dissensus equilibrium of the $N-$agent PCA opinion dynamics (\ref{pca_dynamics}) is locally asymptotically stable.
\begin{theoremEnd}{theorem}
Consider an $N-$agent network where the opinion state of each agent evolved according to (\ref{pca_dynamics}). 
Let $\dot{\bm{v}}_i=\bm{f}(\bm{v})=(\bm{I}-\bm{v}_i\bm{v}_i^\intercal)\bm{C}(\bm{v})\bm{v}_i$. For any $i\in\mathcal{V}$, linearizing $\bm{f}(\bm{v})$ w.r.t. $\bm{v}_i$ leads to $\dot{\bm{v}}_i=\bm{A}_i\bm{v}_i$ where
\begin{align}\label{Ai}
    \bm{A}_i\triangleq\frac{\partial\dot{\bm{v}}_i}{\partial\bm{v}_i}&=\bm{C}+(\bm{v}_i-\bar{\bm{v}})^\intercal\bm{v}_i\bm{I}-(\bm{v}_i-\bar{\bm{v}})\bm{v}_i^\intercal\notag\\
&-\bm{v}_i^\intercal\bm{C}\bm{v}_i\bm{I}-2\bm{v}_i\bm{v}_i^\intercal\bm{C}+2\bm{v}_i\bm{v}_i^\intercal(\bm{v}_i-\bar{\bm{v}})\bm{v}_i^\intercal.
\end{align}
\end{theoremEnd}
\begin{proofEnd}
In this proof, we write $\bm{C}$ for $\bm{C}(\bm{v})$ for simplicity. First we write $\bm{f}(\bm{v})=(\bm{I}-\bm{v}_i\bm{v}_i^\intercal)\bm{C}\bm{v}_i=\bm{C}\bm{v}_i-\bm{v}_i\bm{v}_i^\intercal\bm{C}\bm{v}_i$. 
Then, we obtain
\begin{align}\label{equ0}
&\frac{\partial \bm{f}}{\partial\bm{v}_i}=\frac{\partial (\bm{C}\bm{v}_i)}{\partial \bm{v}_i}-\frac{\partial (\bm{v}_i\bm{v}_i^\intercal\bm{C}\bm{v}_i)}{\partial \bm{v}_i}
\end{align}
In what follows, we first derive $\frac{\partial (\bm{C}\bm{v}_i)}{\partial \bm{v}_i}$ and then we derive  $\frac{\partial (\bm{v}_i\bm{v}_i^\intercal\bm{C}\bm{v}_i)}{\partial \bm{v}_i}$.
\subsection*{\textbf{Derivation of $\frac{\partial (\bm{C}\bm{v}_i)}{\partial \bm{v}_i}$:} }
Note that 
\begin{align}\label{cov_sim}
    &\bm{C}=\sum_{j=1}^N(\bm{v}_j-\bar{\bm{v}})(\bm{v}_j-\bar{\bm{v}})^\intercal=\sum_{j=1}^N\bm{v}_j\bm{v}_j^\intercal-N\bar{\bm{v}}\bar{\bm{v}}^\intercal.
\end{align}
Hence,
\begin{align}\label{eqr0}
\frac{\partial (\bm{C}\bm{v}_i)}{\partial \bm{v}_i}=&-N\frac{\partial }{\partial \bm{v}_i}(\bar{\bm{v}}\bar{\bm{v}}^\intercal\bm{v}_i) +\frac{\partial }{\partial \bm{v}_i}( \sum_{j=1}^N\bm{v}_j\bm{v}_j^\intercal\bm{v}_i).
\end{align}
However, 
\begin{align}\label{eqr1}
    \frac{\partial }{\partial \bm{v}_i} \sum_{j=1}^N\bm{v}_j\bm{v}_j^\intercal\bm{v}_i
    &= \frac{\partial }{\partial \bm{v}_i}(\sum_{j=1,j\ne i}^N\bm{v}_j\bm{v}_j^\intercal\bm{v}_i)+\frac{\partial }{\partial \bm{v}_i}(1\bm{v}_i)\notag\\
    &=\sum_{j=1}^N\bm{v}_j\bm{v}_j^\intercal +\bm{I}-\bm{v}_i\bm{v}_i^\intercal.
\end{align}
On the other hand, since  $N\bar{\bm{v}}\Bar{\bm{v}}^\intercal=\frac1N(\sum_{k=1}^N \bm{v}_k)(\sum_{k=1}^N \bm{v}_k)^\intercal$, then
\begin{align}\label{eqr2}
    &\frac{\partial }{\partial \bm{v}_i}(N\bar{\bm{v}}\bar{\bm{v}}^\intercal\bm{v}_i)\notag\\
    &=\frac1N\frac{\partial }{\partial \bm{v}_i}(\sum_{k=1}^N \bm{v}_k)(\sum_{k=1}^N \bm{v}_k)^\intercal\bm{v}_i\notag\\
    &=\frac1N\frac{\partial }{\partial \bm{v}_i}\left((\sum_{k=1,k\ne i}^N \bm{v}_k)(\sum_{k=1,k\ne i}^N \bm{v}_k)^\intercal\bm{v}_i+(\sum_{k=1,k\ne i}^N \bm{v}_k)
    \right.\notag\\
    &\left.+\bm{v}_i(\sum_{k=1,k\ne i}^N \bm{v}_k)^\intercal\bm{v}_i+\bm{v}_i
    \right)\notag\\
    &=\frac1N(\sum_{k=1,k\ne i}^N \bm{v}_k)(\sum_{k=1,k\ne i}^N \bm{v}_k)^\intercal+\frac1N(\sum_{k=1,k\ne i}^N \bm{v}_k)^\intercal\bm{v}_i\bm{I}\notag\\
    &+\frac1N\bm{v}_i(\sum_{k=1,k\ne i}^N \bm{v}_k)^\intercal+\frac1N\bm{I}\notag\\
    &=\frac1N(\sum_{k=1}^N \bm{v}_k)(\sum_{k=1,k\ne i}^N \bm{v}_k)^\intercal+\frac1N(\sum_{k=1,k\ne i}^N \bm{v}_k)^\intercal\bm{v}_i\bm{I}+\frac1N\bm{I}\notag\\
    &=\frac1N(\sum_{k=1}^N \bm{v}_k)(\sum_{k=1}^N \bm{v}_k)^\intercal -\frac1N(\sum_{k=1}^N \bm{v}_k)\bm{v}_i^\intercal+\frac1N(\sum_{k=1}^N \bm{v}_k)^\intercal\bm{v}_i\bm{I}.
\end{align}
Substituting \eqref{eqr1} and \eqref{eqr2} into \eqref{eqr0}, leads to
\begin{align}\label{equ1}
&\frac{\partial (\bm{C}\bm{v}_i)}{\partial \bm{v}_i}=\sum_{j=1}^N\bm{v}_j\bm{v}_j^\intercal +\bm{I}-\bm{v}_i\bm{v}_i^\intercal
\notag\\
&-\frac{1}{N}\left((\sum_{k=1}^N \bm{v}_k)(\sum_{k=1}^N \bm{v}_k)^\intercal -(\sum_{k=1}^N \bm{v}_k)\bm{v}_i^\intercal+(\sum_{k=1}^N \bm{v}_k)^\intercal\bm{v}_i\bm{I}\right)\notag\\
&=\bm{C}+(\bm{v}_i-\bar{\bm{v}})^\intercal\bm{v}_i\bm{I}-(\bm{v}_i-\bar{\bm{v}})\bm{v}_i^\intercal,
\end{align}
where we used the fact that $\bar{\bm{v}}=\frac1N\sum_{k=1}^N \bm{v}_k$ and thus $(\sum_{k=1}^N \bm{v}_k)(\sum_{k=1}^N \bm{v}_k)^\intercal=N^2\bar{\bm{v}}\bar{\bm{v}}^\intercal$.

\subsection*{\textbf{Derivation of $\frac{\partial (\bm{v}_i\bm{v}_i^\intercal\bm{C}\bm{v}_i)}{\partial \bm{v}_i}$:} }
Using \eqref{cov_sim}, we derive
\begin{align}\label{eqw0}
     \bm{v}_i\bm{v}_i^\intercal\bm{C}\bm{v}_i
     &=\bm{v}_i\bm{v}_i^\intercal\left(\sum_{j=1}^N\bm{v}_j\bm{v}_j^\intercal-N\bar{\bm{v}}\bar{\bm{v}}^\intercal\right)\bm{v}_i\notag\\
     &=\bm{v}_i\bm{v}_i^\intercal(\sum_{j=1}^N\bm{v}_j\bm{v}_j^\intercal)\bm{v}_i-N\bm{v}_i\bm{v}_i^\intercal\bar{\bm{v}}\bar{\bm{v}}^\intercal\bm{v}_i.
\end{align}
Then, we obtain
\begin{align}\label{eqw1}
&\frac{\partial\bm{v}_i\bm{v}_i^\intercal\left(\sum_{j=1}^N\bm{v}_j\bm{v}_j^\intercal\right)\bm{v}_i }{\partial \bm{v}_i}
\notag\\
&=\bm{v}_i^\intercal(\sum_{j=1,j\ne i}^N\bm{v}_j\bm{v}_j^\intercal)\bm{v}_i\bm{I}+2\bm{v}_i\bm{v}_i^\intercal(\sum_{j=1,j\ne i}^N\bm{v}_j\bm{v}_j^\intercal) +\bm{I}\notag\\
&=\bm{v}_i^\intercal(\sum_{j=1}^N\bm{v}_j\bm{v}_j^\intercal)\bm{v}_i\bm{I} +2\bm{v}_i\bm{v}_i^\intercal(\sum_{j=1}^N\bm{v}_j\bm{v}_j^\intercal)-2\bm{v}_i\bm{v}_i^\intercal\notag\\
&=\bm{v}_i^\intercal\bm{C}\bm{v}_i\bm{I}+2\bm{v}_i\bm{v}_i^\intercal\bm{C}+N(\bm{v}_i^\intercal\bar{\bm{v}})^2\bm{I}+2N\bm{v}_i\bm{v}_i^\intercal\bar{\bm{v}}\bar{\bm{v}}^\intercal-2\bm{v}_i\bm{v}_i^\intercal,
\end{align}
where we used the fact that $\sum_{j=1}^N\bm{v}_j\bm{v}_j^\intercal=\bm{C}+N\bar{\bm{v}}\bar{\bm{v}}^\intercal$.
Additionally, using the fact that $\bar{\bm{v}}=\frac1N\sum_{k=1}^N \bm{v}_k\bm{v}_k^\intercal=\frac1N\sum_{k=1,k\ne i}^N \bm{v}_k\bm{v}_k^\intercal+\frac1N\bm{v}_i\bm{v}_i^\intercal$, we can show that 
\begin{align}
    &N\bm{v}_i\bm{v}_i^\intercal\bar{\bm{v}}\bar{\bm{v}}^\intercal\bm{v}_i=\frac1N \bm{v}_i(\sum_{k=1,k\ne i}^N \bm{v}_k)^\intercal\bm{v}_i+\frac1N \bm{v}_i\\
    &+\frac1N \bm{v}_i\bm{v}_i^\intercal(\sum_{k=1,k\ne i}^N \bm{v}_k)(\sum_{k=1,k\ne i}^N \bm{v}_k)^\intercal\bm{v}_i
    +\frac1N \bm{v}_i\bm{v}_i^\intercal(\sum_{k=1,k\ne i}^N \bm{v}_k). \notag
\end{align}
Therefore,
\begin{align}\label{eqs0}
    &\frac{\partial}{\partial\bm{v}_i}N\bm{v}_i\bm{v}_i^\intercal\bar{\bm{v}}\bar{\bm{v}}^\intercal\bm{v}_i
    =\frac1N\frac{\partial}{\partial\bm{v}_i}\left( \bm{v}_i\bm{v}_i^\intercal(\sum_{k=1,k\ne i}^N \bm{v}_k)(\sum_{k=1,k\ne i}^N \bm{v}_k)^\intercal\bm{v}_i\right.\notag\\
    &\left.
    + \bm{v}_i\bm{v}_i^\intercal(\sum_{k=1,k\ne i}^N \bm{v}_k) + \bm{v}_i(\sum_{k=1,k\ne i}^N \bm{v}_k)^\intercal\bm{v}_i+\bm{v}_i\right)\notag\\
    &=\frac1N(\bm{v}_i^\intercal(\sum_{k=1,k\ne i}^N \bm{v}_k)(\sum_{k=1,k\ne i}^N \bm{v}_k)^\intercal\bm{v}_i\bm{I}\notag\\
    &+\bm{v}_i\left(2(\sum_{k=1,k\ne i}^N \bm{v}_k)(\sum_{k=1,k\ne i}^N \bm{v}_k)^\intercal\bm{v}_i\right)^\intercal\notag\\
    &+\frac2N (\bm{v}_i^\intercal(\sum_{k=1,k\ne i}^N \bm{v}_k)\bm{I}+\bm{v}_i(\sum_{k=1,k\ne i}^N \bm{v}_k)^\intercal)+\frac1N\bm{I}\notag\\
    &=\frac1N\bm{v}_i^\intercal(\sum_{k=1,k\ne i}^N \bm{v}_k)(\sum_{k=1,k\ne i}^N \bm{v}_k)^\intercal\bm{v}_i\bm{I}\notag\\
    &+\frac2N\bm{v}_i\bm{v}_i^\intercal(\sum_{k=1,k\ne i}^N \bm{v}_k)(\sum_{k=1,k\ne i}^N \bm{v}_k)^\intercal\notag\\
    &+\frac2N \bm{v}_i^\intercal(\sum_{k=1,k\ne i}^N \bm{v}_k)\bm{I}+\frac2N\bm{v}_i(\sum_{k=1,k\ne i}^N \bm{v}_k)^\intercal+\frac1N\bm{I}.
\end{align}
However, using the fact that $\sum_{k=1}^N \bm{v}_k\bm{v}_k^\intercal=\sum_{k=1,k\ne i}^N \bm{v}_k\bm{v}_k^\intercal+\bm{v}_i\bm{v}_i^\intercal$, we can show that
\begin{align}\label{eqs1}
    &\bm{v}_i^\intercal(\sum_{k=1,k\ne i}^N \bm{v}_k)(\sum_{k=1,k\ne i}^N \bm{v}_k)^\intercal\bm{v}_i\notag\\
    &=\bm{v}_i^\intercal(\sum_{k=1}^N \bm{v}_k)(\sum_{k=1}^N \bm{v}_k)^\intercal\bm{v}_i
    -2(\sum_{k=1}^N \bm{v}_k)^\intercal\bm{v}_i+1\notag\\
    &=N\bm{v}_i^\intercal\bar{\bm{v}}\bar{\bm{v}}^\intercal\bm{v}_i-2N\bar{\bm{v}}^\intercal\bm{v}_i+1,
\end{align}
and
\begin{align}\label{eqs2}
    &\bm{v}_i\bm{v}_i^\intercal(\sum_{k=1,k\ne i}^N \bm{v}_k)(\sum_{k=1,k\ne i}^N \bm{v}_k)^\intercal
    =\bm{v}_i\bm{v}_i^\intercal(\sum_{k=1}^N \bm{v}_k)(\sum_{k=1}^N \bm{v}_k)^\intercal\notag\\
    &-\bm{v}_i(\sum_{k=1}^N \bm{v}_k)^\intercal-\bm{v}_i\bm{v}_i^\intercal(\sum_{k=1}^N \bm{v}_k) \bm{v}_i^\intercal+\bm{v}_i\bm{v}_i^\intercal\notag\\
    &=N^2\bm{v}_i\bm{v}_i^\intercal\bar{\bm{v}}\bar{\bm{v}}^\intercal-N\bm{v}_i\bar{\bm{v}}^\intercal-N\bm{v}_i\bm{v}_i^\intercal\bar{\bm{v}}\bm{v}_i^\intercal+\bm{v}_i\bm{v}_i^\intercal.
\end{align}
Additionally, since $\bm{v}_i^\intercal(\sum_{k=1,k\ne i}^N \bm{v}_k)=\bm{v}_i^\intercal(\sum_{k=1}^N \bm{v}_k)-1$, $\bm{v}_i(\sum_{k=1,k\ne i}^N \bm{v}_k)^\intercal=\bm{v}_i(\sum_{k=1}^N \bm{v}_k)^\intercal-\bm{v}_i\bm{v}_i^\intercal$, and using \eqref{eqs1} and \eqref{eqs2}, we simplify \eqref{eqs0} to be
\begin{align}\label{eqw2}
    &\frac{\partial}{\partial\bm{v}_i}N\bm{v}_i\bm{v}_i^\intercal\bar{\bm{v}}\bar{\bm{v}}^\intercal\bm{v}_i=N(\bm{v}_i^\intercal\bar{\bm{v}})^2\bm{I}+2(N\bm{v}_i\bm{v}_i^\intercal\bar{\bm{v}}\bar{\bm{v}}^\intercal-\bm{v}_i\bm{v}_i^\intercal\bar{\bm{v}}\bm{v}_i^\intercal).
\end{align}
Then, using \eqref{eqw0}, \eqref{eqw1} and \eqref{eqw2}, we obtain
\begin{align}\label{equ2}
\frac{\partial (\bm{v}_i\bm{v}_i^\intercal\bm{C}\bm{v}_i)}{\partial \bm{v}_i}&=\frac{\partial }{\partial \bm{v}_i}\bm{v}_i\bm{v}_i^\intercal\left(\sum_{j=1}^N\bm{v}_j\bm{v}_j^\intercal\right)\bm{v}_i-\frac{\partial}{\partial\bm{v}_i}N\bm{v}_i\bm{v}_i^\intercal\bar{\bm{v}}\bar{\bm{v}}^\intercal\bm{v}_i\notag\\
&=\bm{v}_i^\intercal\bm{C}\bm{v}_i\bm{I}+2\bm{v}_i\bm{v}_i^\intercal\bm{C}-2\bm{v}_i\bm{v}_i^\intercal+2\bm{v}_i\bm{v}_i^\intercal\bar{\bm{v}}\bm{v}_i^\intercal\notag\\
&=\bm{v}_i^\intercal\bm{C}\bm{v}_i\bm{I}+2\bm{v}_i\bm{v}_i^\intercal\bm{C}-2\bm{v}_i\bm{v}_i^\intercal(\bm{v}_i-\bar{\bm{v}})\bm{v}_i^\intercal.
\end{align}
Finally, substituting \eqref{equ1} and \eqref{equ2} into \eqref{equ0} yields the claimed result \eqref{Ai}.
\end{proofEnd}
\begin{theorem}
The matrix $\bm{A}_i$ is negative definite at the dissensus equilibrium $\bm{v}_i=-\bm{v}_j,\quad\forall (i,j)\in\bar{\mathcal{E}}$ where $\bar{\mathcal{E}}\delequal \{(i,j)|i\in\mathcal{V}_1\text{ and }j\in\mathcal{V}_2\}$.
\end{theorem}
\begin{proof}
According to Definition \ref{def_dis}, if the system is at dissensus, then there exist two non-empty sets $\mathcal{V}_1,\mathcal{V}_2$ such that $\mathcal{V}_1\cup\mathcal{V}_2=\{1,\cdots,N\}$ and $\mathcal{V}_1\cup\mathcal{V}_2=\emptyset$. Consider $\bm{s}\in\mathbb{R}^d$ where $\|\bm{s}\|_2=1$ such that $\bm{v}_i=\bm{s}$ for any $i\in\mathcal{V}_1$ and $\bm{v}_j=-\bm{s}$ for any $j\in\mathcal{V}_2$. This implies that $\bm{v}_i\bm{v}_i^\intercal=\bm{s}\bm{s}^\intercal$ for any $i\in\mathcal{V}_1$ and $\bm{v}_j\bm{v}_j^\intercal=(-\bm{s})(-\bm{s})^\intercal=\bm{s}\bm{s}^\intercal$ for any $j\in\mathcal{V}_2$.  Then the averaged opinion state $\bar{\bm{v}}$ can be written as
\begin{align}\label{vabrs}
    \bar{\bm{v}}=\frac1N \sum_{k=1}^N \bm{v}_k=\frac1N(\sum_{i\in\mathcal{V}_1}\bm{v}_i+\sum_{j\in\mathcal{V}_2}\bm{v}_j)=\frac{|\mathcal{V}_1|-|\mathcal{V}_2|}{N} \bm{s},
\end{align}
where $|\mathcal{V}_l|$ represents the cardinality of set $\mathcal{V}_l$, $l=1,2$, and $|\mathcal{V}_1|+|\mathcal{V}_2|=N$, $|\mathcal{V}_1|\geq1$, $|\mathcal{V}_2|\geq1$. The covariance matrix at the dissensus equilibrium becomes
\begin{align}\label{Cs}
    \bm{C}(\bm{v})
    &=\sum_{k=1}^N\bm{v}_k\bm{v}_k^\intercal-N\bar{\bm{v}}\bar{\bm{v}} =\frac{N^2-(|\mathcal{V}_1|-|\mathcal{V}_2|)^2}{N}\bm{s}\bm{s}^\intercal.
\end{align}
 Then, in \eqref{Ai}, substituting \eqref{vabrs} for $\bar{\bm{v}}$, substituting \eqref{Cs} for $\bm{C}$, and substituting $\bm{v}_i=\bm{s}$ for any $i\in\mathcal{V}_1$, yields the linearization matrix $\bm{A}_i$ evaluated at the dissensus equilibrium
\begin{align}
\bm{A}_i
&=-\left(\frac{N^2-(|\mathcal{V}_1|-|\mathcal{V}_2|)^2}{N}-(1-\frac{|\mathcal{V}_1|-|\mathcal{V}_2|}{N})\right)\bm{s}\bm{s}^\intercal\notag\\
&-(\frac{N^2-(|\mathcal{V}_1|-|\mathcal{V}_2|)^2}{N}-(1-\frac{|\mathcal{V}_1|-|\mathcal{V}_2|}{N}))\bm{I}\notag\\
&\triangleq=-m_i(\bm{s}\bm{s}^\intercal+\bm{I}),
\end{align}
where $m_i=(\frac{N^2-(|\mathcal{V}_1|-|\mathcal{V}_2|)^2}{N}-(1-\frac{|\mathcal{V}_1|-|\mathcal{V}_2|}{N}))$ for any $i\in\mathcal{V}_1$.

Similarly, using $\bm{v}_j=-\bm{s}$ for any $j\in\mathcal{V}_1$,  the
linearization matrix $\bm{A}_j$ evaluated at the dissensus equilibrium is
\begin{align}
\bm{A}_j
&=-\left(\frac{N^2-(|\mathcal{V}_1|-|\mathcal{V}_2|)^2}{N}-(1+\frac{|\mathcal{V}_1|-|\mathcal{V}_2|}{N})\right)\bm{s}\bm{s}^\intercal\notag\\
&-(\frac{N^2-(|\mathcal{V}_1|-|\mathcal{V}_2|)^2}{N}-(1+\frac{|\mathcal{V}_1|-|\mathcal{V}_2|}{N}))\bm{I}\notag\\
&\triangleq-m_j(\bm{s}\bm{s}^\intercal+\bm{I}),
\end{align}
where $m_j=(\frac{N^2-(|\mathcal{V}_1|-|\mathcal{V}_2|)^2}{N}-(1+\frac{|\mathcal{V}_1|-|\mathcal{V}_2|}{N}))$ for any $j\in\mathcal{V}_2$.

Since $\mathcal{V}_1\cup\mathcal{V}_2=\{1,\cdots,N\}$ and $\mathcal{V}_1\cap\mathcal{V}_2=\emptyset$, $-(N-2)\leq|\mathcal{V}_1|-|\mathcal{V}_2|\leq N-2$, which leads to for any $i\in\mathcal{V}_1$,
\begin{align}
    m_i&=\frac{N^2-(|\mathcal{V}_1|-|\mathcal{V}_2|)^2}{N}-(1-\frac{|\mathcal{V}_1|-|\mathcal{V}_2|}{N})\notag\\
    &=\frac1N[(N^2-(|\mathcal{V}_1|-|\mathcal{V}_2|)^2)-(N-(|\mathcal{V}_1|-|\mathcal{V}_2|))
    ]\notag\\
    &=\frac1N(N-(|\mathcal{V}_1|-|\mathcal{V}_2|))[(N+(|\mathcal{V}_1|-|\mathcal{V}_2|))-1]\notag\\
    &>0.
\end{align}
Then $\bm{A}_i=-m_i(\bm{s}\bm{s}^\intercal+\bm{I})$ is strictly negative definite for any $i\in\mathcal{V}_1$, since $(\bm{s}\bm{s}^\intercal+\bm{I})$ is strictly positive definite and $m_i>0$.

Following the same procedure, we can also show $m_j=\frac{N^2-(|\mathcal{V}_1|-|\mathcal{V}_2|)^2}{N}-(1+\frac{|\mathcal{V}_1|-|\mathcal{V}_2|}{N})>0$ and $\bm{A}_j=-m_j(\bm{s}\bm{s}^\intercal+\bm{I})$ is strictly negative definite for any $j\in\mathcal{V}_2$.

Therefore, $\bm{A}_k$ is strictly negative definite for any $k\in\mathcal{V}$, which implies that the dissensus equilibrium is asymptotically stable.
\end{proof}
As an example, if $N=2$, then $\bm{A}_1=\bm{A}_2=-(\bm{s}\bm{s}^\intercal+\bm{I})$ which is negative definite. 
\subsection{Lyapunov-based Stable Analysis of the Dissensus Equilibrium for $2-$agent and  $3-$agent Networks.}
In the previous section, we show that the linearized  PCA opinion dynamics \eqref{pca_dynamics} is stable at the dissensus equilibrium. In this section, we instead use Lyapunov stability analysis to study the convergence of the nonlinear PCA opinion dynamics \eqref{pca_dynamics} for the $2-$agent and $3-$agent networks. This analysis reveals the region of attraction of the trajectories of the nonlinear system to the dissensus equilibrium. 
Even for the $2-$agent and $3-$agent cases, the analysis is nontrivial, and thus we leave the generalization to the $N-$agent case to a future work.
\begin{theorem}\label{stable_dis_2}
For a 2-agent system under the PCA dynamics \eqref{pca_dynamics}, the equilibrium $\bm{v}_1=-\bm{v}_2$ is asymptotically stable.
\end{theorem}
\begin{proof}
Define $\beta_{12}=1+\langle\bm{v}_1,\bm{v}_2\rangle$, where 
$\beta_{12}\in[0,2]$ and $\beta_{12}=0$ if and only if $\bm{v}_1=-\bm{v}_2$. 
Consider the Lyapunov candidate function $V:[0,2)\rightarrow \mathbb{R}$ defined by
\begin{align}
    &V=\beta_{12}=1+\langle\bm{v}_1,\bm{v}_2\rangle,
\end{align}
where $V\geq0$ and $V=0$ if and only if $\beta=0$, i.e., $\bm{v}_1=-\bm{v}_2$.
Then, we obtain
\begin{align}\label{uu0}
    \dot{V}&=\langle\bm{v}_2,\dot{\bm{v}}_1\rangle+\langle\bm{v}_1,\dot{\bm{v}}_2\rangle\notag\\
    &=\bm{v}_2^\intercal (\bm{I}-\bm{v}_1\bm{v}_1^\intercal)\bm{C}(\bm{v})\bm{v}_1+\bm{v}_1^\intercal(\bm{I}-\bm{v}_2\bm{v}_2^\intercal)\bm{C}(\bm{v})\bm{v}_2\notag\\
    &=2\bm{v}_1^\intercal\bm{C}(\bm{v})\bm{v}_2-\bm{v}_1^\intercal\bm{v}_2\left(\bm{v}_1^\intercal\bm{C}(\bm{v})\bm{v}_1+\bm{v}_2^\intercal\bm{C}(\bm{v})\bm{v}_2\right),
\end{align}
where $\bm{C}(\bm{v})=\bm{C}(\bm{v}_1,\bm{v}_2)=\frac12(\bm{v}_1-\bm{v}_2)(\bm{v}_1-\bm{v}_2)^\intercal$.
However
\begin{align}\label{uu1}
    &2\bm{v}_1^\intercal\bm{C}(\bm{v})\bm{v}_2\notag\\
    &=(\bm{v}_1+\bm{v}_2)^\intercal\bm{C}(\bm{v})(\bm{v}_1+\bm{v}_2)-(\bm{v}_1^\intercal\bm{C}(\bm{v})\bm{v}_1+\bm{v}_2^\intercal\bm{C}(\bm{v})\bm{v}_2)\notag\\
    &=\|\bm{v}_1+\bm{v}_2\|_{\bm{C}}^2-(\|\bm{v}_1\|_{\bm{C}}^2+\|\bm{v}_2\|_{\bm{C}}^2),
\end{align}
where $\|\bm{x}\|_{\bm{C}}^2=\bm{x}^\intercal\bm{C}(\bm{v})\bm{x}$.
Similarly,
\begin{align}\label{uu2}
    \bm{v}_1^\intercal\bm{v}_2&=\frac{1}{2}(\bm{v}_1+\bm{v}_2)^\intercal(\bm{v}_1+\bm{v}_2)-\frac{1}{2}(\bm{v}_1^\intercal\bm{v}_1+\bm{v}_2^\intercal\bm{v}_2)\notag\\
    &=\frac{1}{2}\|\bm{v}_1+\bm{v}_2\|_2^2-1,
\end{align}
Substituting \eqref{uu1} and \eqref{uu2} into \eqref{uu0}, we obtain
\begin{align}
    \dot{V}&=\|\bm{v}_1+\bm{v}_2\|_{\bm{C}}^2-(\|\bm{v}_1\|_{\bm{C}}^2+\|\bm{v}_2\|_{\bm{C}}^2)\notag\\
    &-(\frac{1}{2}\|\bm{v}_1+\bm{v}_2\|_2^2-1)(\|\bm{v}_1\|_{\bm{C}}^2+\|\bm{v}_2\|_{\bm{C}}^2)\notag\\
    &=\|\bm{v}_1+\bm{v}_2\|_{\bm{C}}^2
    -\frac{1}{2}\|\bm{v}_1+\bm{v}_2\|_2^2(\|\bm{v}_1\|_{\bm{C}}^2+\|\bm{v}_2\|_{\bm{C}}^2).
\end{align}
We know that, for the 2-agent case, $\bm{C}(\bm{v})=\frac12(\bm{v}_1-\bm{v}_2)(\bm{v}_1-\bm{v}_2)^\intercal$. This implies that $ \|\bm{v}_1+\bm{v}_2\|_{\bm{C}}^2=\frac12(\bm{v}_1+\bm{v}_2)^\intercal(\bm{v}_1-\bm{v}_2)(\bm{v}_1-\bm{v}_2)^\intercal(\bm{v}_1+\bm{v}_2)=\frac12(\bm{v}_1^\intercal\bm{v}_1-\bm{v}_1^\intercal\bm{v}_2+\bm{v}_2^\intercal\bm{v}_1-\bm{v}_2^\intercal\bm{v}_2)(\bm{v}_1^\intercal\bm{v}_1+\bm{v}_1^\intercal\bm{v}_2-\bm{v}_2^\intercal\bm{v}_1-\bm{v}_2^\intercal\bm{v}_2)=0$.
This implies  that
\begin{align}
    \dot{V}=
    -\frac{1}{2}\|\bm{v}_1+\bm{v}_2\|_2^2(\|\bm{v}_1\|_{\bm{C}}^2+\|\bm{v}_2\|_{\bm{C}}^2)\notag\leq 0,
\end{align}
where, in the considered domain $D=[0,2)$, $\dot{V}=0$ if and only if $\bm{v}_1=-\bm{v}_2$. Since $V=V(\beta_{12})$ 
is a monotonic function, and bounded below by zero, then the trajectory of the system will stay in a compact sublevel set of the Lyapunov function, implying that domain $D=[0,2)$ is forward invariant. That is, if $0\leq\beta_{12}(0)<2$, then $0\leq\beta_{12}(t)<2$ for all $t\geq0$. Or equivalently, if  $-1\leq\langle\bm{v}_1(0),\bm{v}_2(0)\rangle<1$, then $-1\leq\langle\bm{v}_1(t),\bm{v}_2(t)\rangle<1$ for all $t>0$. Therefore, the equilibrium $\beta_{12}=0$, and hence $\bm{v}_1=-\bm{v}_2$ is asymptotically stable.
\end{proof}

\begin{remark}
In  Theorem~\ref{unstable_con_N} and Theorem~\ref{stable_dis_2}, we have shown that the PCA-based opinion dynamics \eqref{pca_dynamics} yields unstable consensus and stable dissensus, respectively,  for a $2-$agent system on an unsigned graph. This result cannot be achieved by the averaged dissensus algorithms since they require the graph to be signed and weighted, and result in stable dissensus for such a graph \cite{Altafini2013,Ma2018}. The averaged consensus algorithm on the unit sphere, however, can only lead to stable consensus with positive interaction matrix (i.e., unsigned graph), and lead to stable dissensus with sign-symmetric interaction matrix (i.e. signed graph) according to Proposition 1 in \cite{caponigro2015nonlinear}.
\end{remark}
\begin{figure}[htbp]
\centering
\includegraphics[width=6cm]{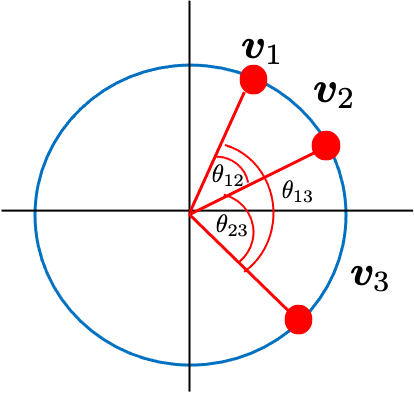}
\caption{Illustration of initial opinion states of a 3-agent system.}
\label{3agents}
\end{figure}
\begin{theorem}\label{stable_dis_3}
Consider a network of three agents where the $2-$dimensional opinion of each agent evolves according to the PCA dynamics \eqref{pca_dynamics}. Suppose that initially $-1\leq\langle\bm{v}_1(0),\bm{v}_3(0)\rangle\leq-a<0$ and $-1\leq\langle\bm{v}_2(0),\bm{v}_3(0)\rangle\leq-b<0$ where $a,b\in(0,1)$. Then the equilibrium $\bm{v}_1=\bm{v}_2=-\bm{v}_3$ is asymptotically stable. 
\end{theorem}
\begin{proof}
Define the set $\bar{\mathcal{E}}\delequal \{(i,j)|i\in\mathcal{V}_1\text{ and }j\in\mathcal{V}_2\}$. For any $(i,j)\in\bar{\mathcal{E}}$, define $\beta_{ij}=1+\langle\bm{v}_i,\bm{v}_j\rangle$ where $\beta_{ij}=0$ if and only if $\bm{v}_i=-\bm{v}_j$.
Consider a Lyapunov candidate function for a general network of $M$ agents defined by
\begin{align}
    &V=\sum_{(i,j)\in\bar{\mathcal{E}}}\beta_{ij}=\sum_{(i,j)\in\bar{\mathcal{E}}}[1+\langle\bm{v}_i,\bm{v}_j\rangle],
\end{align}
where $V\geq0$ and $V=0$ if and only if $\beta_{ij}=0$ for all $(i,j)\in\bar{\mathcal{E}}$, i.e. $\bm{v}_i=-\bm{v}_j,\forall  (i,j)\in\bar{\mathcal{E}}$.
Then, we obtain
\begin{align}\label{vdotxx1}
    &\dot{V}=\sum_{i\in\mathcal{V}_1,j\in\mathcal{V}_2}[\langle\bm{v}_j,\dot{\bm{v}}_i\rangle+\langle\bm{v}_i,\dot{\bm{v}}_j\rangle]=
    \notag\\
    &\sum_{i\in\mathcal{V}_1,j\in\mathcal{V}_2}[\langle\bm{v}_j,(\bm{I}-\bm{v}_i\bm{v}_i^\intercal)\bm{C}(\bm{v})\bm{v}_i\rangle+\langle\bm{v}_i,(\bm{I}-\bm{v}_j\bm{v}_j^\intercal)\bm{C}(\bm{v})\bm{v}_j\rangle]\notag\\
    &=\sum_{i\in\mathcal{V}_1,j\in\mathcal{V}_2}[2\bm{v}_i^\intercal\bm{C}(\bm{v})\bm{v}_j-(\lambda_i+\lambda_j)\bm{v}_i^\intercal\bm{v}_j],
\end{align}
where $\lambda_k=\bm{v}_k^\intercal\bm{C}(\bm{v})\bm{v}_k\in(\lambda_{\text{min}}(\bm{C}(\bm{v})),\lambda_{\text{max}}(\bm{C}(\bm{v})))$, where $\lambda_{\text{min}}(\bm{C}(\bm{v}))$ and $\lambda_{\text{max}}(\bm{C}(\bm{v}))$ are the minimum and maximum eigenvalues of $\bm{C}(\bm{v})$, respectively. 

For a network of three agents, the covariance matrix can be written as $\bm{C}(\bm{v})={\frac13}[(\bm{v}_1-\bm{v}_2)(\bm{v}_1-\bm{v}_2)^\intercal+(\bm{v}_1-\bm{v}_3)(\bm{v}_1-\bm{v}_3)^\intercal+(\bm{v}_2-\bm{v}_3)(\bm{v}_2-\bm{v}_3)^\intercal]$. Hence, we obtain
\begin{align}
    &\lambda_1=\frac13[(1-\bm{v}_1^\intercal\bm{v}_2)^2+(1-\bm{v}_1^\intercal\bm{v}_3)^2+(\bm{v}_1^\intercal\bm{v}_2-\bm{v}_1^\intercal\bm{v}_3)^2],\label{Lm1}\\
    &\lambda_2=\frac13[(1-\bm{v}_1^\intercal\bm{v}_2)^2+(\bm{v}_1^\intercal\bm{v}_2-\bm{v}_2^\intercal\bm{v}_3)^2+(1-\bm{v}_2^\intercal\bm{v}_3)^2],\label{Lm2}\\
    &\lambda_3=\frac13[(\bm{v}_1^\intercal\bm{v}_3-\bm{v}_2^\intercal\bm{v}_3)^2+(1-\bm{v}_1^\intercal\bm{v}_3)^2+(1-\bm{v}_2^\intercal\bm{v}_3)^2].\label{Lm3}
\end{align}
Suppose without loss of generality that $\mathcal{V}_1=\{1,2\}$ and $\mathcal{V}_2=\{3\}$, i.e. $\bar{\mathcal{E}}=\{(1,3),(2,3)\}$.
Then, we derive $\sum_{(i,j)\in\bar{\mathcal{E}}}
     \bm{v}_i^\intercal\bm{C}(\bm{v})\bm{v}_j
     =\bm{v}_1^\intercal\bm{C}(\bm{v})\bm{v}_3+\bm{v}_2^\intercal\bm{C}(\bm{v})\bm{v}_3=-\frac13(1-\bm{v}_1^\intercal\bm{v}_3)^2-\frac13(1-\bm{v}_2^\intercal\bm{v}_3)^2-\frac13(1-\bm{v}_2^\intercal\bm{v}_3)(\bm{v}_1^\intercal\bm{v}_2-\bm{v}_1^\intercal\bm{v}_3)-\frac13(1-\bm{v}_1^\intercal\bm{v}_3)(\bm{v}_1^\intercal\bm{v}_2-\bm{v}_2^\intercal\bm{v}_3)$.
On the other hand, we derive $-\sum_{(i,j)\in\bar{\mathcal{E}}}
     (\lambda_i+\lambda_j)\bm{v}_i^\intercal\bm{v}_j
     =-(\lambda_1+\lambda_3)\bm{v}_1^\intercal\bm{v}_3-(\lambda_2+\lambda_3)\bm{v}_2^\intercal\bm{v}_3$. 
Therefore
\begin{align}
    &\dot{V}=-W+Q,
\end{align}
where
\begin{align}
    W&=\frac23\left[(1-\bm{v}_1^\intercal\bm{v}_3)^2+(1-\bm{v}_2^\intercal\bm{v}_3)^2+(1-\bm{v}_2^\intercal\bm{v}_3)(\bm{v}_1^\intercal\bm{v}_2-\bm{v}_1^\intercal\bm{v}_3)\right.\notag\\
     &\left.+(1-\bm{v}_1^\intercal\bm{v}_3)(\bm{v}_1^\intercal\bm{v}_2-\bm{v}_2^\intercal\bm{v}_3)\right]\notag\\
     &=\frac23\left[(2-\beta_{13})^2+(2-\beta_{23})^2+(2-\beta_{23})(1-\beta_{13}+\bm{v}_1^\intercal\bm{v}_2)\right.\notag\\
     &\left.+(2-\beta_{13})(1-\beta_{23}+\bm{v}_1^\intercal\bm{v}_2)\right]
\end{align}
and 
\begin{align}
    &Q
    =-(\lambda_1+\lambda_3)\bm{v}_1^\intercal\bm{v}_3-(\lambda_2+\lambda_3)\bm{v}_2^\intercal\bm{v}_3\notag\\
    &=(\lambda_1+\lambda_3)(1-\beta_{13})+(\lambda_2+\lambda_3)(1-\beta_{23}),
\end{align}
where $\lambda_1\geq0$, $\lambda_2\geq0$, and $\lambda_3\geq0$ are as given by \eqref{Lm1}-\eqref{Lm3}.
Note that  $\dot{V}=0$ if and only if $\bm{v}_1=\bm{v}_2=-\bm{v}_3$, i.e. $(\beta_{13},\beta_{23})=(0,0)$.
If $-1\leq\langle\bm{v}_1(0),\bm{v}_3(0)\rangle\leq-a<0$ and $-1\leq\langle\bm{v}_2(0),\bm{v}_3(0)\rangle\leq-b<0$ where $a,b\in(0,1)$, then this implies that $\beta_{13}(0)\in(0,1-a)$ and  $\beta_{23}(0)\in(0,1-b)$. For $2-$dimensional opinion states, assume without loss of generality that the three agents are initially distributed as shown in Fig.~\ref{3agents}. Then this implies that $\langle\bm{v}_1(0),\bm{v}_2(0)\rangle=\cos(\theta_{13}(0)-\theta_{23}(0))=\cos(\cos^{-1}(\beta_{13}(0)-1)-\cos^{-1}(\beta_{23}(0)-1))\geq0$. Hence, $W(0)\geq0$ and $Q(0)\geq0$. 

Since $V(0)=\beta_{13}(0)+\beta_{23}(0)\leq2-(a+b)$, then for any $a,b\in(0,1)$, we can show that $\dot{V}(0)<0$. For example, if $a=b=0.0001$, then $\langle\bm{v}_1(0),\bm{v}_2(0)\rangle\approx1$ and $\dot{V}(0)=-2.664$, and if $a=0.8$, $b=0.0001$, then $\langle\bm{v}_1(0),\bm{v}_2(0)\rangle\approx0.6$ and $\dot{V}(0)=-1.1519$.
Since $V(\beta_{13},\beta_{23})=\beta_{13}+\beta_{23}$ is a monotonic function  and bounded below by zero, then the trajectory of the system will stay in a compact sublevel set of the Lyapunov function, implying that
$-1\leq\langle\bm{v}_1(t),\bm{v}_3(t)\rangle\leq-a<0$ and $-1\leq\langle\bm{v}_2(t),\bm{v}_3(t)\rangle\leq-b<0$ for all $t>0$. Therefore, the origin $(\beta_{13},\beta_{23})=(0,0)$ 
 is asymptotically stable.
\end{proof}
Note that, if $0<\langle\bm{v}_1(0),\bm{v}_3(0)\rangle\leq a<1$ and $-1\leq\langle\bm{v}_2(0),\bm{v}_3(0)\rangle\leq b<0$, then using the same proof we can show that the PCA system \eqref{pca_dynamics} will converge to the equilibrium $\bm{v}_2=-\bm{v}_1=-\bm{v}_3$. Similarly,  if $-1\leq\langle\bm{v}_1(0),\bm{v}_3(0)\rangle\leq -a<0$ and $0<\langle\bm{v}_2(0),\bm{v}_3(0)\rangle\leq b<1$ where $a,b\in(0,1)$, then the PCA system \eqref{pca_dynamics} will converge to the equilibrium $\bm{v}_1=-\bm{v}_2=-\bm{v}_3$.

\begin{remark}
We have shown that the dissensus states of an N-agent system are locally asymptotically stable through linearization. So far, we are able to construct Lyapunov function for $N\le 3$. 
Lyapunov stability analysis for $N\ge 4$ will be investigated in future work. 
\end{remark}

\section{Simulation Result}\label{simulation}

In this section, we present a simulation result in $\mathbb{R}^2$ that demonstrates a stable dissensus equilibria of the PCA-based opinion dynamics \eqref{pca_dynamics} in a multi-agent system. 

\begin{figure}[htbp]
\centering
\includegraphics[width=8.5cm]{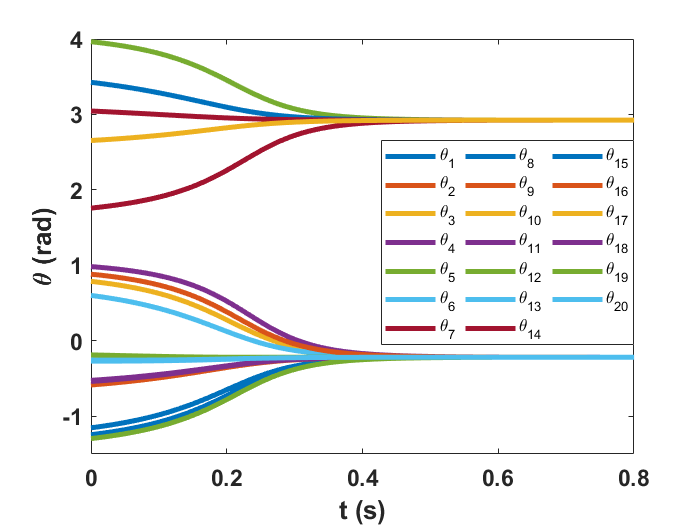}
\caption{Stable antipodal equilibrium for a 20-agent system in $\mathbb{R}^2$}
\label{20_anti}
\end{figure}

In $\mathbb{R}^2$, any unit vector $\bm{v}_i$ can be represented by $\bm{ v}_i=[\cos\theta_i,\sin\theta_i]^\intercal$. Suppose the opinion states initially located at a non-equilibrium position on the unit circle in $\mathbb{R}^2$. Fig.\ref{20_anti} shows the evolution of $\theta_i$ in a 20-agent system which demonstrates the transition of opinion states from non-equilibrium to dissensus equilibrium. While evolving on the unit sphere. In this example, based on the initial opinions, the 20 agents split into 2 sub-groups $\mathcal{V}_1,\mathcal{V}_2$.  The opinion states in each subgroup converge to a consensus state. Additionally, the consensus state of one subgroup is the opposite of the consensus state of the other subgroup. The emergence of the two subgroup is clearly indicated in Fig.\ref{20_anti} when  $|\theta_i-\theta_j|=\pi$ for any $i\in\mathcal{V}_1,j\in\mathcal{V}_2$, which occurred at time t=0.8s.

\section{Conclusion}\label{conclusion}
In this paper, we propose novel nonlinear modeling of opinion dynamics based on the Oja PCA flow. We discovered that a stable dissensus equilibrium can be achieved by the PCA-based dynamics with a varying covariance matrix regardless of the initial opinion states of the agents. However, if the covariance matrix is fixed, then neither consensus nor dissensus can be guaranteed for all initial opinion states. In the future, we will extend the Lyapunov-based stability analysis to the general $N-$agent complete and incomplete networks. 

\section*{acknowledgements}
Z. Zhang, S. Al-Abri and F. Zhang were supported by ONR grants N00014-19-1-2556 and N00014-19-1-2266; AFOSR grant FA9550-19-1-0283; NSF grants  CNS-1828678, S\&AS-1849228 and GCR-1934836; NRL grants N00173-17-1-G001 and N00173-19-P-1412; and NOAA grant NA16NOS0120028.

\section{Proofs}\label{proofs}
\printProofs

\bibliographystyle{IEEEtran}       
\bibliography{ref}
\end{document}